\documentclass[journal=jacsat,manuscript=article]{achemso}

\usepackage{chemformula} 
\usepackage[T1]{fontenc} 

\usepackage{bm}
\usepackage{braket}
\renewcommand{\v}[1]{\bm{\mathrm{#1}}}

\usepackage{csquotes}
\MakeOuterQuote{"}

\author{Ruikai Wu}
\affiliation[BigPharma]
{Max-Born-Institute for Non-Linear optics, Max-Born Strasse 2A, 12489 Berlin, Germany}

\author{Deepika Gill}
\affiliation[BigPharma]
{Max-Born-Institute for Non-Linear optics, Max-Born Strasse 2A, 12489 Berlin, Germany}

\author{Sangeeta Sharma}
\affiliation[BigPharma]
{Max-Born-Institute for Non-Linear optics, Max-Born Strasse 2A, 12489 Berlin, Germany}
\affiliation[BigBigPharma]
{Institute for theoretical solid-state physics, Freie Universit\"at Berlin, Arnimallee 14, 14195 Berlin, Germany.}

\author{Sam Shallcross}
\email{shallcross@mbi-berlin.de}
\affiliation[BigPharma]
{Max-Born-Institute for Non-Linear optics, Max-Born Strasse 2A, 12489 Berlin, Germany}

\title[An \textsf{achemso} demo]
{Ultrafast ghost Hall states in a 2d altermagnet}

\abbreviations{IR,NMR,UV}

\begin{document}

\begin{tocentry}

\includegraphics[width=0.5\textwidth]{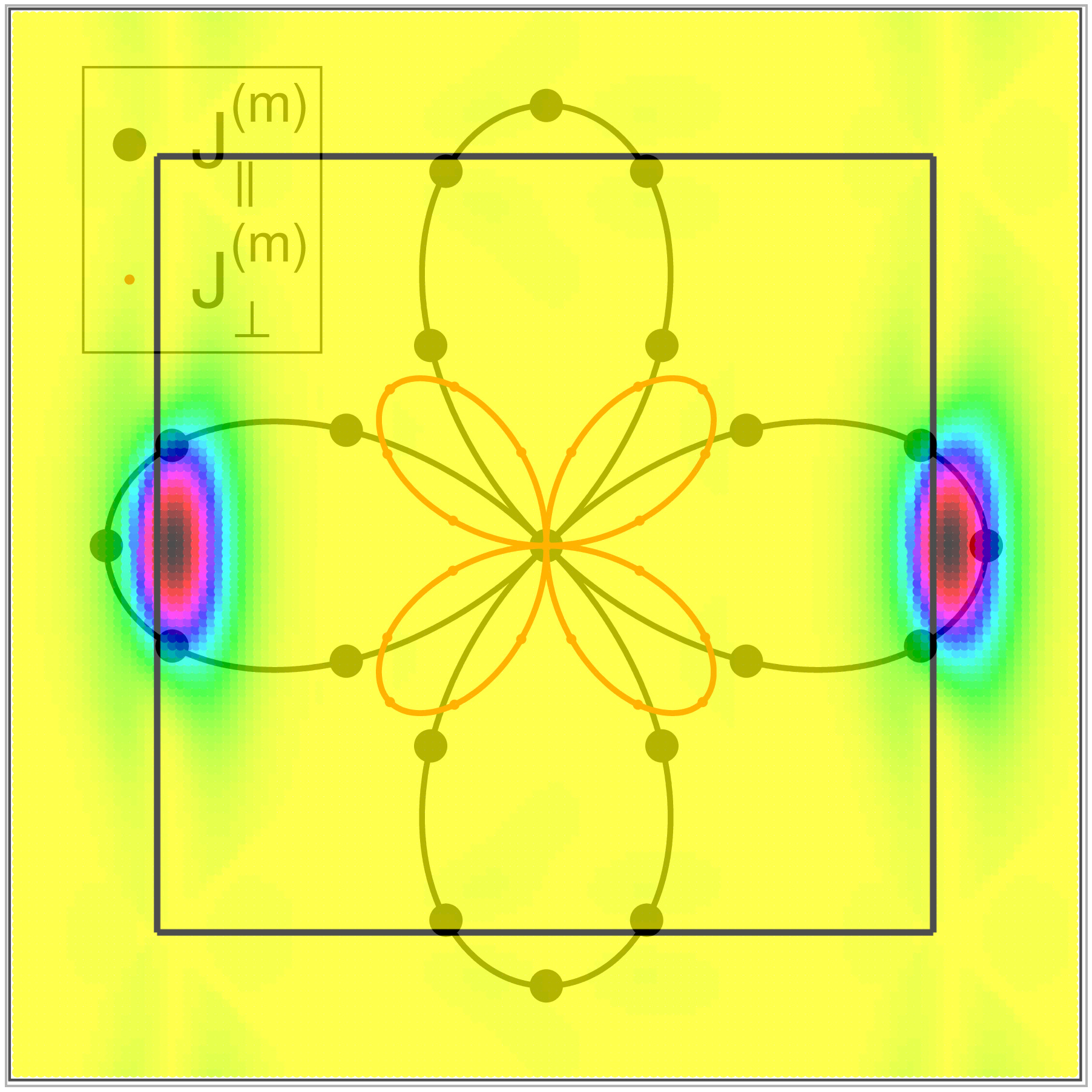}

\end{tocentry}

\begin{abstract}
Two-dimensional materials that exhibit optically active spin and valley degrees of freedom represent one of the most fascinating -- and potentially most technologically useful -- platforms for the ultrafast interaction of light and matter. Here we show, via the example of Cr$_2$SO, that two dimensional altermagnets host valley states controllable by femtosecond laser light: linearly polarized light pulses excite charge at one of two inequivalent valleys, with which valley charge is excited at determined by the polarization vector direction. This underpins a rich spin and valley physics including: (i) valleytronics -- the generation of nearly 100\% spin polarized valley currents, as well as (ii) a "ghost Hall" effect -- the ultrafast creation of states in which spin and charge currents are orthogonal without invoking Hall physics. Our findings establish 2d altermagents as a platform providing a new route for the control of spin- and charge currents at ultrafast times.
\end{abstract}

\section{Introduction}

Altermagnets represent a new form of magnetic matter, combining in one material the two most characteristic features of ferromagnets and anti-ferromagnets: exchange split bands and fully compensated magnetic order\cite{bai_altermagnetism_2024}. Underpinning this novel electronic state is a k-dependent exchange splitting, alternating in sign throughout the Brillouin zone according to a characteristic d- g- or even i-wave symmetry. This in turn generates a remarkable physics of transport\cite{gonzalez-hernandez_efficient_2021,
zhang_electrical_2025,bai_efficient_2023,
zhang_simultaneous_2024,ullah_giant_2024}, including such phenomena as a "pseudo-Hall" effect in which a electric field directed along a particular crystalline direction generates a perpendicular spin current.

A second, dramatically different, approach to harnessing the solid state to generate and control currents of spin and charge is via ultrafast excitation through femtosecond laser pulses\cite{schiffrin_optical-field-induced_2013,sharma_thz_2023}.
In this strongly non-equilibrium regime both semi-conducting as well as metallic states can be addressed as platforms for current based physics, and indeed it is in two dimensional (2d) semi-conductors such as the dichalcogenides that a particularly rich physics of ultrafast light control is found.
Selective light activation\cite{xiao_coupled_2012} of discrete local extrema in the conduction and valence band edges ("valleys") allows the creation and control of both spin- and valley charge states, as well as their associated currents\cite{sharma_thz_2023,
sharma_giant_2023,sharma_light-shaping_2023}, with a wealth of reported ultrafast phenomena
\cite{mak_control_2012,xiao_nonlinear_2015,
langer_lightwave_2018,berghauser_inverted_2018,
ishii_optical_2019,
silva_all-optical_2022,sharma_valley_2022,
langer_lightwave_2018}. The recent discovery of a class of two dimensional altermagnets thus raises the possibility of ultrafast light control over valley states in the novel context of electronic bands imprinted with an alternating exchange splitting.

Here we report that linearly polarized laser pulses can controllably excite the two distinct valleys of the 2d altermagnet Cr$_2$SO, a finding that holds both for single cycle few femtosecond pulses as well as in the multi-cycle regime.
The $d$-wave exchange field generates, however, a valleytronics quite distinct from that found in 2d dichalcogenides, with femtosecond laser pulses able to create both massive spin polarized currents, but also "ghost Hall" states featuring ultrafast excitation of perpendicular spin and charge currents. Our work points towards a rich valley physics in the 2d semi-conducting altermagnets, with profound control over spin currents, establishing these materials as a promising platform for ultrafast spin- and valleytronics.

\section{Ultrafast valleytronics}

To explore the physics of light-matter coupling we employ a dual approach consisting of (i) a Wannier parametrized time-dependent tight-binding scheme (TD-TB) and (ii) state-of-the-art time-dependent density functional theory (TD-DFT) as implemented in the Elk code\cite{elk}. While the ground state band structures are very similar, the former method allows dynamics  only via the time dependent occupation numbers, as the band structure is fixed to that of the ground state. In contrast, in TD-DDT the full many-body density represents the dynamical object. The latter approach, considerably more numerically demanding, we employ as a quality check of results obtained by TD-TB. 

\begin{figure}[t!]
\begin{center}
\includegraphics[width=0.99\textwidth]{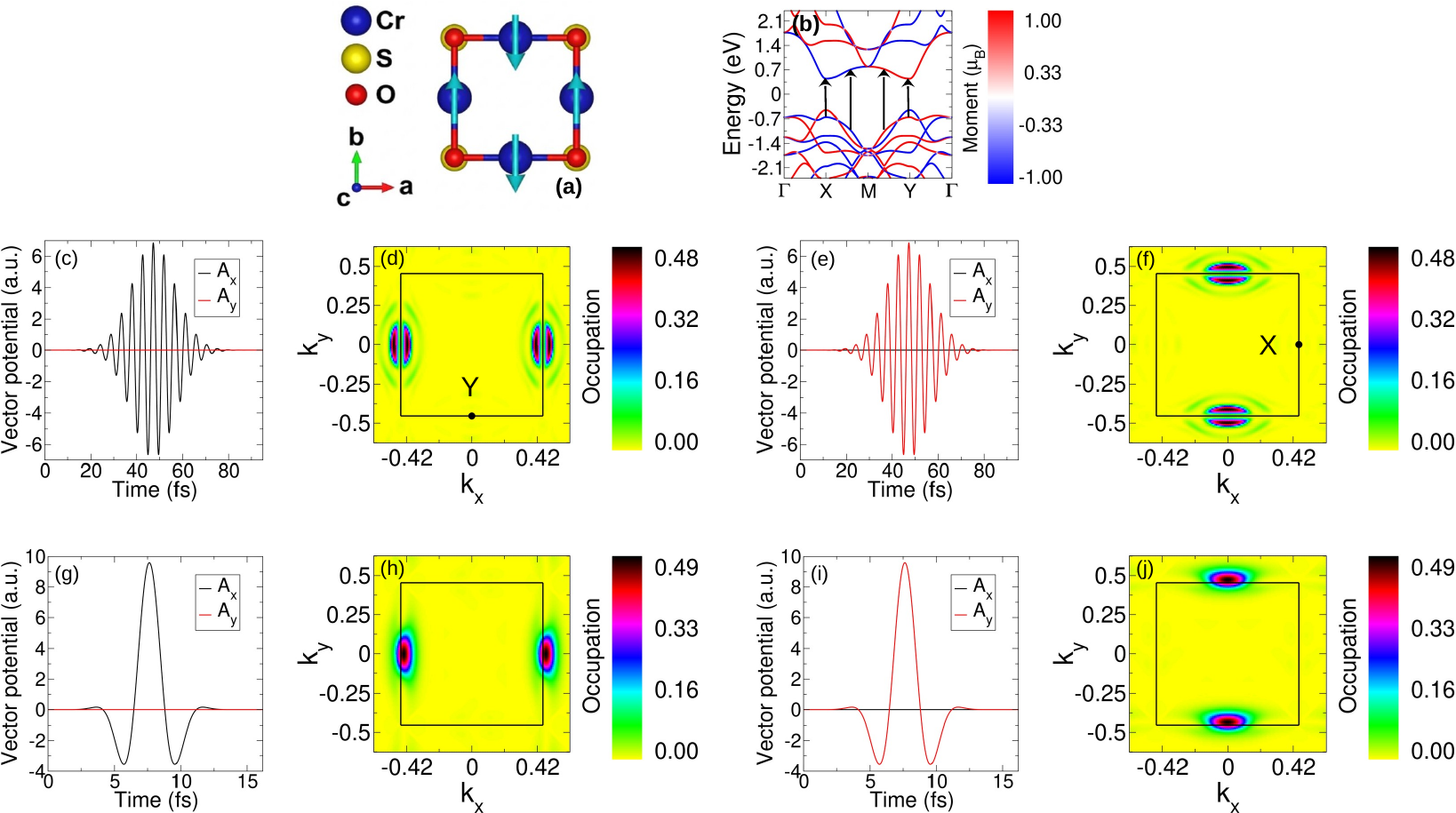}
\caption{{\it Light control over valley excitation}. (a) The lattice structure and (b) band structure of Cr$_2$SO, with arrows indicating possible laser induced transitions at the valleys. Gap tuned (0.88~eV) $x$-polarized light, vector potential shown in panel (c), generates excitation exclusively at the X valley of Cr$_2$SO, as revealed in the momentum resolved charge excitation, panel (d). Similarly, $y$-polarized light generates excitation exclusively at the Y valley, panels (e,f). This holds both for multi-cycle pulses but also in the single cycle limit, panels (g-j), in which the valley charge excitation can be seen to be displaced off the valley centre.
}
\label{fig1}
\end{center}
\end{figure}

As a materials platform we employ Cr$_2$SO\cite{guo_hidden_2026}, chosen as a representative example of the class of two dimensional $d$-wave square ("Lieb") lattice altermagnets
; the lattice structure of this material can be found in Fig.~\ref{fig1}(a). The corresponding band structure, calculated in the LDA+U scheme with $U=3.5$~eV on the Cr sites, presents characteristic valley structures at the high symmetry X and Y points, with the $d$-wave exchange field symmetry of this material ensuring that these valleys have opposite polarization of the bands. Further details of our methodology, and numerical parameters the calculations, can be found in Supporting Information.

Linearly polarized light pulses whose polarization vectors are perpendicular to the X and Y special points couple exclusively to the X and Y valleys respectively, as may be seen in the momentum resolved charge excitation,  Fig.~\ref{fig1}(c-f). Here we have employed gap tuned laser light (0.89~eV) with a full width half maxima (FWHM) of 22.3~fs and a fluence of 0.26~mJ/cm$^2$. Reducing the pulse duration to the single cycle regime, Fig.~\ref{fig1}(g-j), yields again charge excitation only at the X and Y valleys. Thus both for long duration multi-cycle waveforms, as well as the strong field single cycle limit, a light-valley selection rule holds.

Comparison of Fig.~\ref{fig1}(c-f) with Fig.~\ref{fig1}(g-j) reveals a distinct lowering of the local $C_2$ valley symmetry in the charge excitation: while the momentum resolved valley excitation obeys $C_2$ symmetry in the case of the multi-cycle pulse, this symmetry is broken in the single cycle pulse. The valley-current response of these two pulses will thus markedly differ, as symmetry lowering of the charge excitation inevitably leads to a net valley current as contributions from $\pm k$ states no longer cancel.

Such generation of charge current by few cycle pulses has in fact been observed experientially in the case of graphene, with few cycle pulses generating highly anisotropic excitation of the K and K$^\ast$ valleys. In the case of Cr$_2$SO, however, the X and Y valleys possess opposite spin, and so endowing these valleys with light-induced current opens up a rich possibilities for light wave control over spin current. 

\section{Ultrafast ghost-Hall states}

\begin{figure}[t!]
\begin{center}
\includegraphics[width=0.99\textwidth]{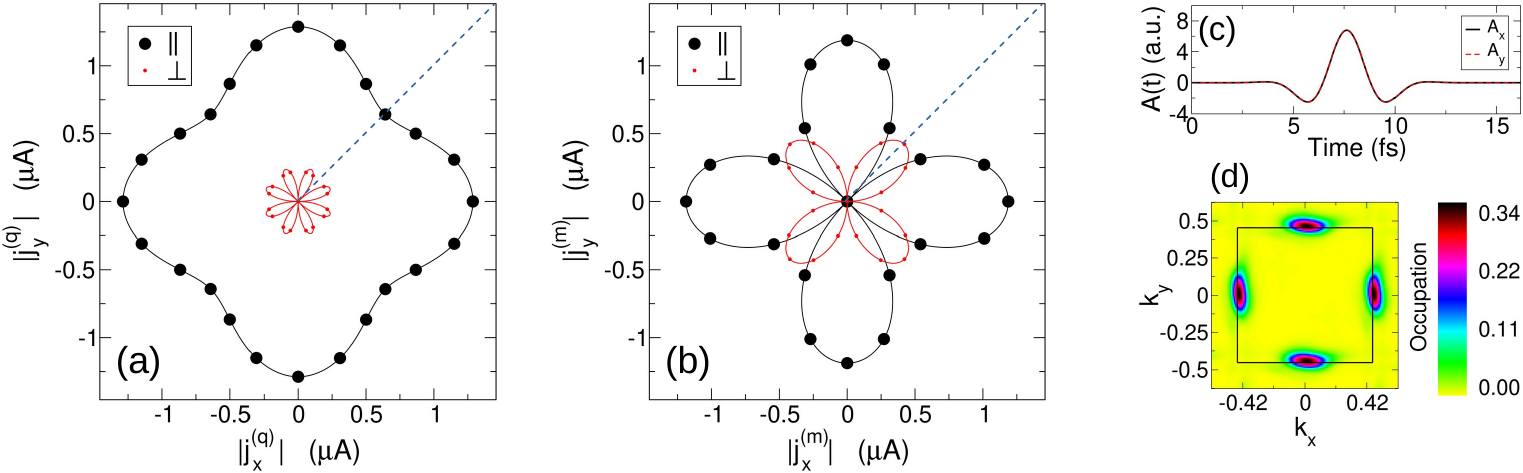}
\caption{{\it Light induced spin polarized currents and ultrafast "ghost" Hall states}. Linearly polarized laser light induces spin and charge currents both parallel and perpendicular to the polarization vector angle $\theta_L$. (a) In a polar plot over $\theta_L$ the charge current is seen to possess nodes in the perpendicular component ($\perp$), but none in the parallel component ($\parallel$) which is finite for all $\theta_L$. (b) In dramatic contrast the spin current presents nodes in both parallel and perpendicular components. Comparison of (a) and (b) reveals that for $\theta_L = n \pi/2$ ($n$ integer), i.e. 0$^\circ$, 90$^\circ$, 180$^\circ$, and 270$^\circ$, spin and charge currents are nearly equal and parallel to the polarization vector of the laser pulse, while for $\theta_L = \pi/4 + n \pi/2$, i.e. 45$^\circ$, 135$^\circ$, 225$^\circ$, 315$^\circ$, charge and spin currents are respectively parallel and perpendicular to the polarization vector of the laser pulse. The former represents the case of a light induced near fully spin polarized current, with the latter an ultrafast "ghost" Hall state. 
}
\label{fig2}
\end{center}
\end{figure}

To explore this we take the light pulse of Fig.~\ref{fig1}(e) and continuously rotate the polarization vector through 360$^\circ$.
For each of these cases we then examine the total light induced charge and spin currents, focussing on the intra-band component that represents the "useful" steady state post-excitation residual current.
Breaking the charge ($\v j^{(q)}$) and spin ($\v j^{(m)}$) current into components parallel and perpendicular to the applied polarization vector of the laser pulse, we see that the charge current always has a finite parallel component, with the perpendicular component vanishing at angles $n \pi/4$ with $n$ an integer, see Fig.~\ref{fig2}(a) in which we present the magnitude $|j^{(q)}_\parallel|$ and $|j^{(q)}_\perp|$ as a function of the polarization angle.
The spin current, presented in identical fashion in Fig.~\ref{fig2}(b), exhibits dramatically different behaviour: the parallel component has nodes at $\pi/4 + n\pi/2$ while the perpendicular component possesses notes at $n \pi/2$. 

Two distinct regimes of polarization angle behaviour can thus be identified. For polarization angles close to the $x$- and $y$-directions an ultrafast light pulse induces a valley polarized excitation endowed with a nearly 100\% polarized spin current. In contrast when the polarization vector aligns with one of the diagonals, one finds a finite charge current parallel to the polarization vector of the light pulse, with a spin current perpendicular to it. For the case of the polarization vector aligned at 45$^\circ$ the corresponding vector potential and momentum resolved excitation are shown in Fig.~\ref{fig2}(c) and (d) respectively. The charge excitation of what we denote a "ghost" Hall state is thus seen to be valley unpolarized, a finding in consonance with the selection rule coupling $x$- and $y$-polarized light to the X and Y valleys; an light pulse composed equally of $x$ and $y$ components will excite charge equally at the X and Y valleys.

\begin{figure}[t!]
\begin{center}
\includegraphics[width=0.7\textwidth]{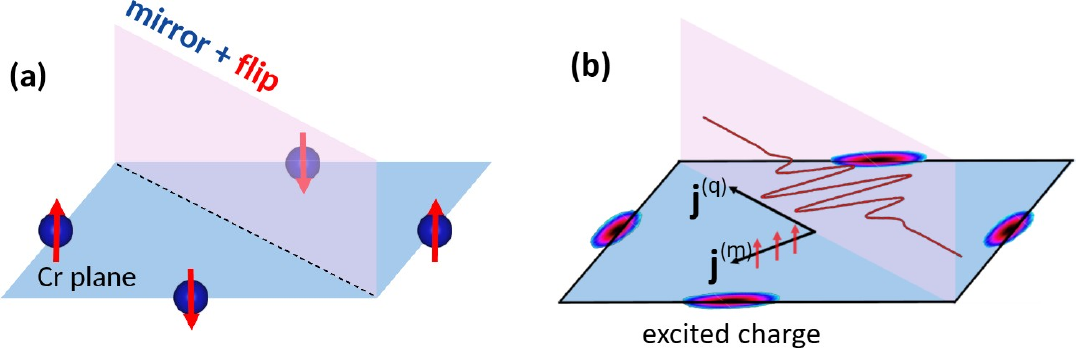}
\caption{{\it Symmetry origin of ghost Hall effect}. (a) The Cr atoms in the structure of Cr$_2$SO illustrating the reflection and spin flip symmetry operation. (b) An ultrafast pulse whose polarization vector is aligned along the 45$^\circ$ direction may, according to this symmetry, excite only charge current parallel to the 45$^\circ$ direction is invariant, and only spin current perpendicular to the 45$^\circ$ direction.
}
\label{fig3}
\end{center}
\end{figure}

The occurrence of this ultrafast "ghost" Hall state for laser light aligned parallel to one of the unit cell diagonals is, in fact, guaranteed by the magnetic point group. In the situation in which the polarization vector of the Hamiltonian aligned with a diagonal both lattice and laser commute with the combined operation of "spin flip" ($\theta$) and reflection about a diagonal ($\sigma_d$). Only charge current parallel to a diagonal is invariant under this combined operation while, in contrast, only a spin current {\it perpendicular} to a diagonal is invariant under the combined operation, as both the reflection and spin flip operations switch the sign of the current vector. These symmetry operations on the charge and spin current vectors are illustrated in Fig.~\ref{fig3}.

This "ghost" Hall state appears to be a direct ultrafast counterpart to the pseudo-Hall effect reported in metallic altermagnets, in which an electric field applied in certain crystallographic directions generates perpendicular charge and spin currents. The context of a valley active semi-conducting altermagnet yields key differences: (i) ultrafast few femtosecond generation times and (ii) very large currents due to significant symmetry lowering of the valley charge excitation for ultrashort laser pulses.

\section{Selection rule for light-altermagnet coupling}

At the high symmetry X and Y points the coupling of linearly polarized light to charge excitation can be motivated by point group symmetry arguments applied to the Bloch states. However, in the strong field regime the large vector potential amplitude drives significant intra-band evolution of crystal momenta, which can be of the order of the Brillouin zone. To establish a light-matter selection rule for this regime, i.e for the dynamics described in the previous section, we thus require a selection rule for the entire band manifold, not only at the special points.

\begin{figure}[t!]
\begin{center}
\includegraphics[width=0.99\textwidth]{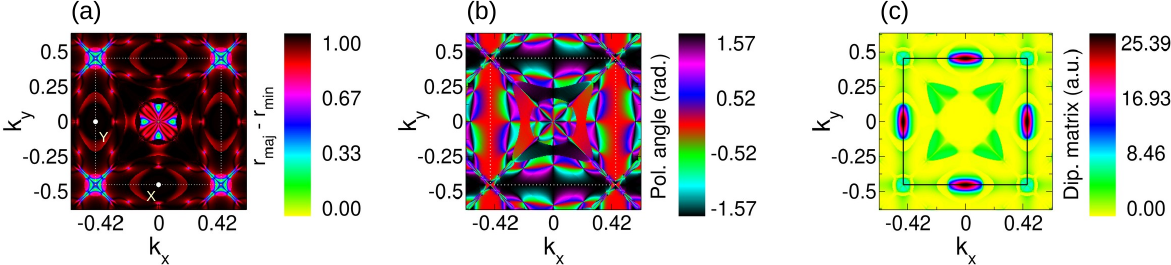}
\caption{{\it Global selection rule underpinning light control of valley excitation in Cr$_2$SO}. (a) The linearly versus circularly polarized character of the light-form generating maximal excitation, expressed as a difference between the major and minor axis of the generally elliptical polarization. This parameter takes of values of 1 for linearly polarized light and 0 for circularly polarized light. Evidently a large domain of linearly polarized coupling exists near the X and Y valleys. (b) The polarization angle of the linearly polarized that maximizes transition, revealing that with large domains centred on the X and Y valleys, excitation is by $x$/$y$-polarized light. (c) Deviations within these regions from this global selection rule are associated with very low dipole matrix elements.
}
\label{fig4}
\end{center}
\end{figure}

To that end we follow a recent work revealing a saddle-point selection rule in graphene, and consider excitation by a general harmonic form $\v A(t) = \v u \cos\omega t + \v v \sin\omega t$. To calculate the transmission $T$ in the linear response regime (i.e. Fermi's golden rule) we break the harmonic form into excitation and de-excitation components $\v A(t) = \v A^+ e^{i\omega t} + \v A^- e^{-i\omega t}$ thus finding for the excitation transmission

\begin{equation}
T = 2\pi \left| \braket{\psi_c|\nabla_{\v k}H_0(\v k).\v A^+|\psi_v} \right|^2 \delta(\epsilon_v(\v k)-\epsilon_c(\v k) - \omega)
\end{equation}
where $H_0(\v k).\v A^+$ represents the excitation part of the light-matter coupling in the velocity gauge, with $H_0(\v k)$ the ground-state Hamiltonian. The harmonic form  $\v A(t)$ yields $\v A^+ = (\v u - i \v v)/2$ and so maximizing the probability of excitation at crystal momentum $\v k$ implies maximizing the expression

\begin{equation}
|\braket{\psi_c|\nabla_{\v k}H_0(\v k)|\psi_v}.(\v u - i \v v)|^2.
\label{maxterm}
\end{equation}
The vectors $\v u_{max}(\v k)$ and $\v u_{max}(\v k)$ that achieve this are, by immediate inspection, seen to be

\begin{eqnarray}
\v u_{max}(\v k) = \text{Re} \braket{\psi_c|\nabla_{\v k}H_0(\v k)|\psi_v} \\
\v v_{max}(\v k) = \text{Im} \braket{\psi_c|\nabla_{\v k}H_0(\v k)|\psi_v}
\end{eqnarray}
These can be calculated directly from the ground state tight-binding Hamiltonian, thus providing the desired map over momentum space of the light pulse form that maximizes excitation i.e. the k-dependent selection rule. 

We now apply this formalism to Cr$_2$SO, considering excitation from the highest energy valence band to lowest energy conduction band (which is always spin preserving as indicated by the arrows in Fig.~\ref{fig4}a). To present this momentum resolved light-matter response it is convenient to express the general elliptically polarized light-form described by the vectors $\v u_{max}(\v k)$ and $\v v_{max}(\v k)$ in terms of three derived quantities: (i) the difference of the major and minor axis, $r_{maj}-r_{min}$, which takes on a value of $+1$ for linearly polarized light and $0$ for circularly polarized light (with the generic case of elliptically polarized light a number between 0 and 1); (ii) the angle of the polarization vector; and (iii) and pulse helicity (which takes on the values of $\pm 1$). 

As can be seen in Fig.~\ref{fig4}(a), not only at the X and Y points, but in a wide region of momentum space near these high symmetry points, $r_{maj}-r_{min} = 1$ and so linearly polarized light maximizes the light induced excitation even for large vector potential amplitude. The corresponding polarization vector, Fig.~\ref{fig4}(c), is aligned along the $x$/$y$-direction in the regions surrounding the X/Y point. In fact, in regions where this does not hold the corresponding dipole matrix element, Fig.~\ref{fig4}(d), is seen to be very small. This establishes a global selection rule described by a tiling of the Brillouin zone into regions in which $x$-polarized and $y$-polarized light drive excitation, with the X and Y special points at the centres of these tiles.

\section{Current strength and pulse duration}

\begin{figure}[t!]
\begin{center}
\includegraphics[width=0.99\textwidth]{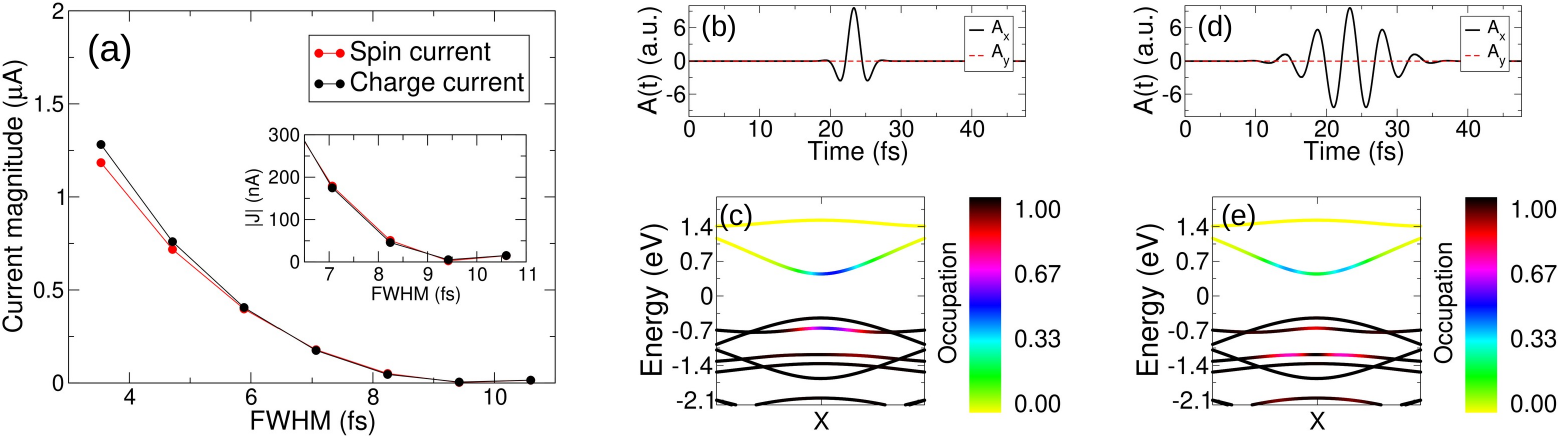}
\caption{{\it Increase of laser pulse full width half maxima (FWHM) acts to reduce the magnitude of light-induced spin and charge currents}. While increasing pulse duration increases energy deposited into the altermagnet, the current magnitude shows a dramatic decrease from $\mu$A to nA scale, panel (a). This is driven by the reduced anisotropy in $\pm A_x$ with increased pulse duration, with concomitant reduction in the anisotropy in occupation of $\pm k$ valley states, compare panels (b,c) and (d,e). As $\pm$ anisotropy in valley occupation underpins the creation of a current carrying state (as $\pm$ states contribute oppositely to the overall current), increase in pulse duration reduces the magnitude of the light-induced currents.
}
\label{fig5}
\end{center}
\end{figure}

Strong field light matter interaction can generate significant displacement of charge from a valley centre, and such excitations -- due to the strong anisotropy in occupation of states with local valley momenta $\pm k$ -- can generate massive currents,
much greater than can be achieved in the perturbative transport regime. The existence of a spin current valley physics in 2d semi-conducting altermagnets thus represents a route towards massive ultrafast generation of both highly spin polarized currents, but also "ghost" Hall currents.

In the single cycle limit of linearly polarized light the charge and spin currents induced by light in Cr$_2$SO are on the $\mu$A scale, see Fig.~\ref{fig2}. Increasing the pulse duration reduces the anisotropy in $\pm A$, in turn reducing the anisotropy in the occupation of $\pm k$ valley states and reducing the current: increasing the pulse full width half maximum into the multi-cycle regime dramatically reduces the current from $\mu$A to nA scale, Fig.~\ref{fig5}. Laser pulse vector potentials and corresponding "fatband" plots of the band occupation post-pulse illustrate this change, with a 3.2~fs single cycle pulse showing very strong anisotropy in both $A_x(t)$ and the valley occupation, panels (b-c), which is significantly reduced at 10.2~fs pulse duration, panels (d-e).

\section{Discussion}

Ultrafast light control over valley charge and current excitations has been demonstrated in the 2d altermagnet Cr$_2$SO. The square lattice and $d$-wave spin splitting render this altermagnet version of valleytronic physics strikingly different to that well established in materials such as graphene and WSe$_2$. In particular, the spin current response to laser light is much richer, with both (i) nearly 100\% spin-polarized light-induced currents and (ii) perpendicular light-induced charge and spin currents found as the polarization angle of the laser pulse is varied.

This physics is underpinned by a selection rule coupling $x$/$y$-linearly polarized light to the X/Y valleys of this material, which we show holds not only at these high symmetry points, which follows from symmetry arguments, but globally throughout the Brillouin zone. Strong field ultrafast light pulses, that can induced massive intra-band evolution of crystal momentum, are thus governed the same valley selection rule as weak amplitude long duration light pulses, opening the way to a rich ultrafast valletronics in these materials. These findings establish "Lieb lattice" 2d altermagnets as ideal platforms for light control over valley and spin physics, promising potential new routes towards valletronics.

\begin{acknowledgement}

Sharma would like to thank SAW for funding through project K612/2024, and Wu would like to thank DFG for funding through project-ID  SH-498-9/1. Sharma and Shallcross would like to thank the Leibniz Professorin Program (SAW P118/2021). The authors acknowledge the North-German Supercomputing Alliance (HLRN) for providing HPC resources that have contributed to the research results reported in this paper.

\end{acknowledgement}

\begin{suppinfo}

Supporting information is available:
\begin{itemize}
  \item Filename: SI.pdf -- this document contains supplemental calculations and associated discussion referred to in the text but not elaborated upon.
\end{itemize}

\end{suppinfo}


\begin{mcitethebibliography}{21}
\providecommand*\natexlab[1]{#1}
\providecommand*\mciteSetBstSublistMode[1]{}
\providecommand*\mciteSetBstMaxWidthForm[2]{}
\providecommand*\mciteBstWouldAddEndPuncttrue
  {\def\EndOfBibitem{\unskip.}}
\providecommand*\mciteBstWouldAddEndPunctfalse
  {\let\EndOfBibitem\relax}
\providecommand*\mciteSetBstMidEndSepPunct[3]{}
\providecommand*\mciteSetBstSublistLabelBeginEnd[3]{}
\providecommand*\EndOfBibitem{}
\mciteSetBstSublistMode{f}
\mciteSetBstMaxWidthForm{subitem}{(\alph{mcitesubitemcount})}
\mciteSetBstSublistLabelBeginEnd
  {\mcitemaxwidthsubitemform\space}
  {\relax}
  {\relax}

\bibitem[Bai \latin{et~al.}(2024)Bai, Feng, Liu, Šmejkal, Mokrousov, and
  Yao]{bai_altermagnetism_2024}
Bai,~L.; Feng,~W.; Liu,~S.; Šmejkal,~L.; Mokrousov,~Y.; Yao,~Y.
  Altermagnetism: {Exploring} {New} {Frontiers} in {Magnetism} and
  {Spintronics}. \emph{Advanced Functional Materials} \textbf{2024}, \emph{34},
  2409327\relax
\mciteBstWouldAddEndPuncttrue
\mciteSetBstMidEndSepPunct{\mcitedefaultmidpunct}
{\mcitedefaultendpunct}{\mcitedefaultseppunct}\relax
\EndOfBibitem
\bibitem[González-Hernández \latin{et~al.}(2021)González-Hernández,
  Šmejkal, Výborný, Yahagi, Sinova, Jungwirth, and
  Železný]{gonzalez-hernandez_efficient_2021}
González-Hernández,~R.; Šmejkal,~L.; Výborný,~K.; Yahagi,~Y.; Sinova,~J.;
  Jungwirth,~T.; Železný,~J. Efficient {Electrical} {Spin} {Splitter} {Based}
  on {Nonrelativistic} {Collinear} {Antiferromagnetism}. \emph{Physical Review
  Letters} \textbf{2021}, \emph{126}, 127701, Publisher: American Physical
  Society\relax
\mciteBstWouldAddEndPuncttrue
\mciteSetBstMidEndSepPunct{\mcitedefaultmidpunct}
{\mcitedefaultendpunct}{\mcitedefaultseppunct}\relax
\EndOfBibitem
\bibitem[Zhang \latin{et~al.}(2025)Zhang, Bai, Dai, Han, Chen, Liang, Cao,
  Zhang, Wang, Zhu, Pan, and Song]{zhang_electrical_2025}
Zhang,~Y.; Bai,~H.; Dai,~J.; Han,~L.; Chen,~C.; Liang,~S.; Cao,~Y.; Zhang,~Y.;
  Wang,~Q.; Zhu,~W.; Pan,~F.; Song,~C. Electrical manipulation of spin
  splitting torque in altermagnetic {RuO2}. \emph{Nature Communications}
  \textbf{2025}, \emph{16}, 5646, Publisher: Nature Publishing Group\relax
\mciteBstWouldAddEndPuncttrue
\mciteSetBstMidEndSepPunct{\mcitedefaultmidpunct}
{\mcitedefaultendpunct}{\mcitedefaultseppunct}\relax
\EndOfBibitem
\bibitem[Bai \latin{et~al.}(2023)Bai, Zhang, Zhou, Chen, Wan, Han, Zhu, Liang,
  Su, Han, Pan, and Song]{bai_efficient_2023}
Bai,~H.; Zhang,~Y.; Zhou,~Y.; Chen,~P.; Wan,~C.; Han,~L.; Zhu,~W.; Liang,~S.;
  Su,~Y.; Han,~X.; Pan,~F.; Song,~C. Efficient {Spin}-to-{Charge} {Conversion}
  via {Altermagnetic} {Spin} {Splitting} {Effect} in {Antiferromagnet}
  \$\{{\textbackslash}mathrm\{{RuO}\}\}\_\{2\}\$. \emph{Physical Review
  Letters} \textbf{2023}, \emph{130}, 216701, Publisher: American Physical
  Society\relax
\mciteBstWouldAddEndPuncttrue
\mciteSetBstMidEndSepPunct{\mcitedefaultmidpunct}
{\mcitedefaultendpunct}{\mcitedefaultseppunct}\relax
\EndOfBibitem
\bibitem[Zhang \latin{et~al.}(2024)Zhang, Bai, Han, Chen, Zhou, Back, Pan,
  Wang, and Song]{zhang_simultaneous_2024}
Zhang,~Y.; Bai,~H.; Han,~L.; Chen,~C.; Zhou,~Y.; Back,~C.~H.; Pan,~F.;
  Wang,~Y.; Song,~C. Simultaneous {High} {Charge}-{Spin} {Conversion}
  {Efficiency} and {Large} {Spin} {Diffusion} {Length} in {Altermagnetic}
  {RuO2}. \emph{Advanced Functional Materials} \textbf{2024}, \emph{34},
  2313332, \_eprint:
  https://advanced.onlinelibrary.wiley.com/doi/pdf/10.1002/adfm.202313332\relax
\mciteBstWouldAddEndPuncttrue
\mciteSetBstMidEndSepPunct{\mcitedefaultmidpunct}
{\mcitedefaultendpunct}{\mcitedefaultseppunct}\relax
\EndOfBibitem
\bibitem[Ullah \latin{et~al.}(2024)Ullah, Bezzerga, and Hong]{ullah_giant_2024}
Ullah,~A.; Bezzerga,~D.; Hong,~J. Giant spin seebeck effect with highly
  polarized spin current generation and piezoelectricity in flexible {V2SeTeO}
  altermagnet at room temperature. \emph{Materials Today Physics}
  \textbf{2024}, \emph{47}, 101539\relax
\mciteBstWouldAddEndPuncttrue
\mciteSetBstMidEndSepPunct{\mcitedefaultmidpunct}
{\mcitedefaultendpunct}{\mcitedefaultseppunct}\relax
\EndOfBibitem
\bibitem[Schiffrin \latin{et~al.}(2013)Schiffrin, Paasch-Colberg, Karpowicz,
  Apalkov, Gerster, Mühlbrandt, Korbman, Reichert, Schultze, Holzner, Barth,
  Kienberger, Ernstorfer, Yakovlev, Stockman, and
  Krausz]{schiffrin_optical-field-induced_2013}
Schiffrin,~A. \latin{et~al.}  Optical-field-induced current in dielectrics.
  \emph{Nature} \textbf{2013}, \emph{493}, 70--74\relax
\mciteBstWouldAddEndPuncttrue
\mciteSetBstMidEndSepPunct{\mcitedefaultmidpunct}
{\mcitedefaultendpunct}{\mcitedefaultseppunct}\relax
\EndOfBibitem
\bibitem[Sharma \latin{et~al.}(2023)Sharma, Elliott, and
  Shallcross]{sharma_thz_2023}
Sharma,~S.; Elliott,~P.; Shallcross,~S. {THz} induced giant spin and valley
  currents. \emph{Science Advances} \textbf{2023}, \emph{9}, eadf3673,
  Publisher: American Association for the Advancement of Science\relax
\mciteBstWouldAddEndPuncttrue
\mciteSetBstMidEndSepPunct{\mcitedefaultmidpunct}
{\mcitedefaultendpunct}{\mcitedefaultseppunct}\relax
\EndOfBibitem
\bibitem[Xiao \latin{et~al.}(2012)Xiao, Liu, Feng, Xu, and
  Yao]{xiao_coupled_2012}
Xiao,~D.; Liu,~G.-B.; Feng,~W.; Xu,~X.; Yao,~W. Coupled {Spin} and {Valley}
  {Physics} in {Monolayers} of {MoS}$_2$ and {Other} {Group}-{VI}
  {Dichalcogenides}. \emph{Physical Review Letters} \textbf{2012}, \emph{108},
  196802\relax
\mciteBstWouldAddEndPuncttrue
\mciteSetBstMidEndSepPunct{\mcitedefaultmidpunct}
{\mcitedefaultendpunct}{\mcitedefaultseppunct}\relax
\EndOfBibitem
\bibitem[Sharma \latin{et~al.}(2023)Sharma, Gill, and
  Shallcross]{sharma_giant_2023}
Sharma,~S.; Gill,~D.; Shallcross,~S. Giant and {Controllable} {Valley}
  {Currents} in {Graphene} by {Double} {Pumped} {THz} {Light}. \emph{Nano
  Letters} \textbf{2023}, \emph{23}, 10305--10310\relax
\mciteBstWouldAddEndPuncttrue
\mciteSetBstMidEndSepPunct{\mcitedefaultmidpunct}
{\mcitedefaultendpunct}{\mcitedefaultseppunct}\relax
\EndOfBibitem
\bibitem[Sharma \latin{et~al.}(2023)Sharma, Dewhurst, and
  Shallcross]{sharma_light-shaping_2023}
Sharma,~S.; Dewhurst,~J.~K.; Shallcross,~S. Light-{Shaping} of {Valley}
  {States}. \emph{Nano Letters} \textbf{2023}, \emph{23}, 11533--11539\relax
\mciteBstWouldAddEndPuncttrue
\mciteSetBstMidEndSepPunct{\mcitedefaultmidpunct}
{\mcitedefaultendpunct}{\mcitedefaultseppunct}\relax
\EndOfBibitem
\bibitem[Mak \latin{et~al.}(2012)Mak, He, Shan, and Heinz]{mak_control_2012}
Mak,~K.~F.; He,~K.; Shan,~J.; Heinz,~T.~F. Control of valley polarization in
  monolayer {MoS}$_2$ by optical helicity. \emph{Nature Nanotechnology}
  \textbf{2012}, \emph{7}, 494--498\relax
\mciteBstWouldAddEndPuncttrue
\mciteSetBstMidEndSepPunct{\mcitedefaultmidpunct}
{\mcitedefaultendpunct}{\mcitedefaultseppunct}\relax
\EndOfBibitem
\bibitem[Xiao \latin{et~al.}(2015)Xiao, Ye, Wang, Zhu, Wang, and
  Zhang]{xiao_nonlinear_2015}
Xiao,~J.; Ye,~Z.; Wang,~Y.; Zhu,~H.; Wang,~Y.; Zhang,~X. Nonlinear optical
  selection rule based on valley-exciton locking in monolayer {WS}$_2$.
  \emph{Light: Science \& Applications} \textbf{2015}, \emph{4},
  e366--e366\relax
\mciteBstWouldAddEndPuncttrue
\mciteSetBstMidEndSepPunct{\mcitedefaultmidpunct}
{\mcitedefaultendpunct}{\mcitedefaultseppunct}\relax
\EndOfBibitem
\bibitem[Langer \latin{et~al.}(2018)Langer, Schmid, Schlauderer, Gmitra,
  Fabian, Nagler, Schüller, Korn, Hawkins, Steiner, Huttner, Koch, Kira, and
  Huber]{langer_lightwave_2018}
Langer,~F.; Schmid,~C.~P.; Schlauderer,~S.; Gmitra,~M.; Fabian,~J.; Nagler,~P.;
  Schüller,~C.; Korn,~T.; Hawkins,~P.~G.; Steiner,~J.~T.; Huttner,~U.;
  Koch,~S.~W.; Kira,~M.; Huber,~R. Lightwave valleytronics in a monolayer of
  tungsten diselenide. \emph{Nature} \textbf{2018}, \emph{557}, 76--80\relax
\mciteBstWouldAddEndPuncttrue
\mciteSetBstMidEndSepPunct{\mcitedefaultmidpunct}
{\mcitedefaultendpunct}{\mcitedefaultseppunct}\relax
\EndOfBibitem
\bibitem[Bergh\"auser \latin{et~al.}(2018)Bergh\"auser, Bernal-Villamil,
  Schmidt, Schneider, Niehues, Erhart, Michaelis~de Vasconcellos, Bratschitsch,
  Knorr, and Malic]{berghauser_inverted_2018}
Bergh\"auser,~G.; Bernal-Villamil,~I.; Schmidt,~R.; Schneider,~R.; Niehues,~I.;
  Erhart,~P.; Michaelis~de Vasconcellos,~S.; Bratschitsch,~R.; Knorr,~A.;
  Malic,~E. Inverted valley polarization in optically excited transition metal
  dichalcogenides. \emph{Nature Communications} \textbf{2018}, \emph{9}, 971,
  Number: 1 Publisher: Nature Publishing Group\relax
\mciteBstWouldAddEndPuncttrue
\mciteSetBstMidEndSepPunct{\mcitedefaultmidpunct}
{\mcitedefaultendpunct}{\mcitedefaultseppunct}\relax
\EndOfBibitem
\bibitem[Ishii \latin{et~al.}(2019)Ishii, Yokoshi, and
  Ishihara]{ishii_optical_2019}
Ishii,~S.; Yokoshi,~N.; Ishihara,~H. Optical selection rule of monolayer
  transition metal dichalcogenide by an optical vortex. \emph{Journal of
  Physics: Conference Series} \textbf{2019}, \emph{1220}, 012056, Publisher:
  IOP Publishing\relax
\mciteBstWouldAddEndPuncttrue
\mciteSetBstMidEndSepPunct{\mcitedefaultmidpunct}
{\mcitedefaultendpunct}{\mcitedefaultseppunct}\relax
\EndOfBibitem
\bibitem[Silva \latin{et~al.}(2022)Silva, Silva, Ivanov, Ivanov, Ivanov,
  Jim{\'e}nez-Gal{\'a}n, and Jim{\'e}nez-Gal{\'a}n]{silva_all-optical_2022}
Silva,~R. E.~F.; Silva,~R. E.~F.; Ivanov,~M.; Ivanov,~M.; Ivanov,~M.;
  Jim{\'e}nez-Gal{\'a}n,~{\'A}.; Jim{\'e}nez-Gal{\'a}n,~{\'A}. All-optical
  valley switch and clock of electronic dephasing. \emph{Optics Express}
  \textbf{2022}, \emph{30}, 30347--30355, Publisher: Optica Publishing
  Group\relax
\mciteBstWouldAddEndPuncttrue
\mciteSetBstMidEndSepPunct{\mcitedefaultmidpunct}
{\mcitedefaultendpunct}{\mcitedefaultseppunct}\relax
\EndOfBibitem
\bibitem[Sharma \latin{et~al.}(2022)Sharma, Elliott, and
  Shallcross]{sharma_valley_2022}
Sharma,~S.; Elliott,~P.; Shallcross,~S. Valley control by linearly polarized
  laser pulses: example of {WSe}$_{\textrm{2}}$. \emph{Optica} \textbf{2022},
  \emph{9}, 947--952, Publisher: Optica Publishing Group\relax
\mciteBstWouldAddEndPuncttrue
\mciteSetBstMidEndSepPunct{\mcitedefaultmidpunct}
{\mcitedefaultendpunct}{\mcitedefaultseppunct}\relax
\EndOfBibitem
\bibitem[Dewhurst \latin{et~al.}(Jan. 14 {\bf 2018})Dewhurst, Sharma, and
  et~al.]{elk}
Dewhurst,~J.~K.; Sharma,~S.; et~al., Jan. 14 {\bf 2018};
  \url{elk.sourceforge.net}\relax
\mciteBstWouldAddEndPuncttrue
\mciteSetBstMidEndSepPunct{\mcitedefaultmidpunct}
{\mcitedefaultendpunct}{\mcitedefaultseppunct}\relax
\EndOfBibitem
\bibitem[Guo(2026)]{guo_hidden_2026}
Guo,~S.-D. Hidden altermagnetism. \emph{Frontiers of Physics} \textbf{2026},
  \emph{21}, 25201, arXiv:2411.13795 [cond-mat]\relax
\mciteBstWouldAddEndPuncttrue
\mciteSetBstMidEndSepPunct{\mcitedefaultmidpunct}
{\mcitedefaultendpunct}{\mcitedefaultseppunct}\relax
\EndOfBibitem
\end{mcitethebibliography}
\providecommand{\latin}[1]{#1}
\makeatletter
\providecommand{\doi}
  {\begingroup\let\do\@makeother\dospecials
  \catcode`\{=1 \catcode`\}=2 \doi@aux}
\providecommand{\doi@aux}[1]{\endgroup\texttt{#1}}
\makeatother
\providecommand*\mcitethebibliography{\thebibliography}
\csname @ifundefined\endcsname{endmcitethebibliography}
  {\let\endmcitethebibliography\endthebibliography}{}

\end{document}